\documentclass[conference]{IEEEtran}
\usepackage{cite}

\ifCLASSINFOpdf
   \usepackage[pdftex]{graphicx}
   \DeclareGraphicsExtensions{.pdf,.jpeg,.png}
   \usepackage{adjustbox}
\else
\fi
\begin{document}
%
\title{Time Accuracy Analysis of Post-Mediation Packet-Switched Charging Data Records\\ for Urban Mobility Applications}

\author{\IEEEauthorblockN{Oscar F. Peredo}
\IEEEauthorblockA{Telef\'onica I+D, Chile\\
}
\and
\IEEEauthorblockN{Romain Deschamps}
\IEEEauthorblockA{Telef\'onica I+D, Chile\\
}
}


%


\maketitle

\begin{abstract}
Telecommunication data is being used increasingly in urban mobility applications around the world. 
Despite its ubiquity and usefulness, technical difficulties arise when using Packet-Switched 
Charging Data Records (CDR), since its main purpose was not intended for this kind of applications. 
Due to its particular nature, a trade-off must be considered between accessibility and time accuracy when using
this data. On the one hand, to obtain highly accurate timestamps, huge amounts of network-level 
CDR must be extracted and stored. This task is very difficult and expensive since 
highly critical network node applications can be compromised in the data extraction and storage.  
On the other hand, post-mediation CDR can be easily accessed since no network node application is involved in
its analysis. The pay-off is in the lower accurate timestamps recorded, since several aggregations and filtering
is performed in previous steps of the charging pipelines.  
In this work, a detailed description of the timestamp error problem using post-mediation CDR is presented, 
together with a methodology to analyze
error time series collected in each network cell. 
\end{abstract}


%
\IEEEpeerreviewmaketitle

\section{Introduction}

\begin{figure}[!h]
\centering
\includegraphics[width=0.9\columnwidth]{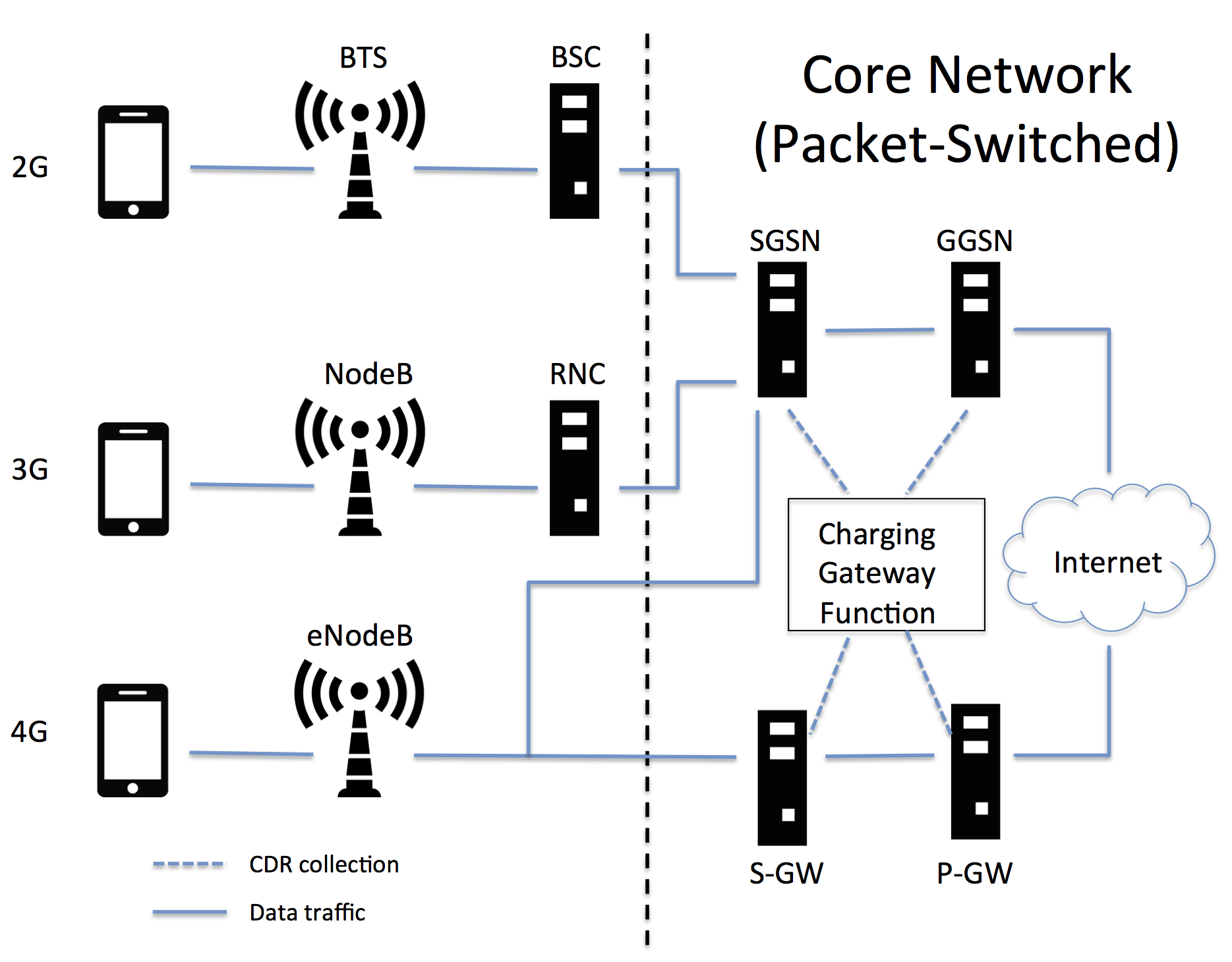}
\caption{Core network simplified diagram \cite{ericsson}.}
\label{fig:network}
\end{figure}

In the context of mobile telecommunications, \emph{Charging Data Records} (CDR) are 
one of the most essential datasets generated and processed by a service provider.
Circuit and packet switched events (Voice, SMS, IP, VoIP and similar) 
are registered by many components of the core network, generating a wide range of CDR types for different purposes.
In terms of volume, the amount of CDR generated by packet switched events (related with Internet traffic and services)
is several orders of magnitude larger than the circuit switched counterparts (Voice, SMS).
According to \cite{ericsson}, a simplified diagram of the core network components and connections involved in the packet switched traffic 
is depicted in Fig.~\ref{fig:network}.
Internal gateway nodes (\emph{SGSN}, \emph{GGSN}, \emph{S-GW} and \emph{P-GW})
handle internet traffic between each subscriber device and applications server, and also generate charging data handled by special network functions (\emph{Charging Gateway Function}).
As the generated events can be geo-located using the latitude/longitude of the corresponding network cell (\emph{BTS}, \emph{NodeB} and \emph{eNodeB}),
packet switched events can be potentially used in the context of urban mobility applications. 

Two main difficulties arise in order to use these events in urban mobility project.
The first one is related with the technical complexity involved in the data generation, which requires a considerable amount of \emph{know-how} of the daily 
operation, business processes and network behaviour. Unlike circuit switched events, the internet traffic charging processes 
apply filters and aggregations in several CDR at different nodes of the network (\cite{korhonen2003} Sect.~10.2 and \cite{cox2014introduction} Sect~13.5). The main problem with this behaviour is
the addition of potential inaccuracies in the event timestamps, which is a critical feature in urban mobility.  
The second one is related with the technical complexity involved in the data storage and access, since the amount of daily data is too large. Big Data 
persistent storages must be used to keep all CDR for posterior analysis in this case. Additionally, it could happen that 
the network/operations departments of the mobile operator can refuse to deploy automatic Extraction-Transformation-Load (ETL) processes in the core 
network nodes. The reason of this rejection is the possibility that the quality of the service to be deteriorated as consequence
of the ETL processes resource consumption.

\begin{figure}[!h]
\centering
\includegraphics[width=0.8\columnwidth]{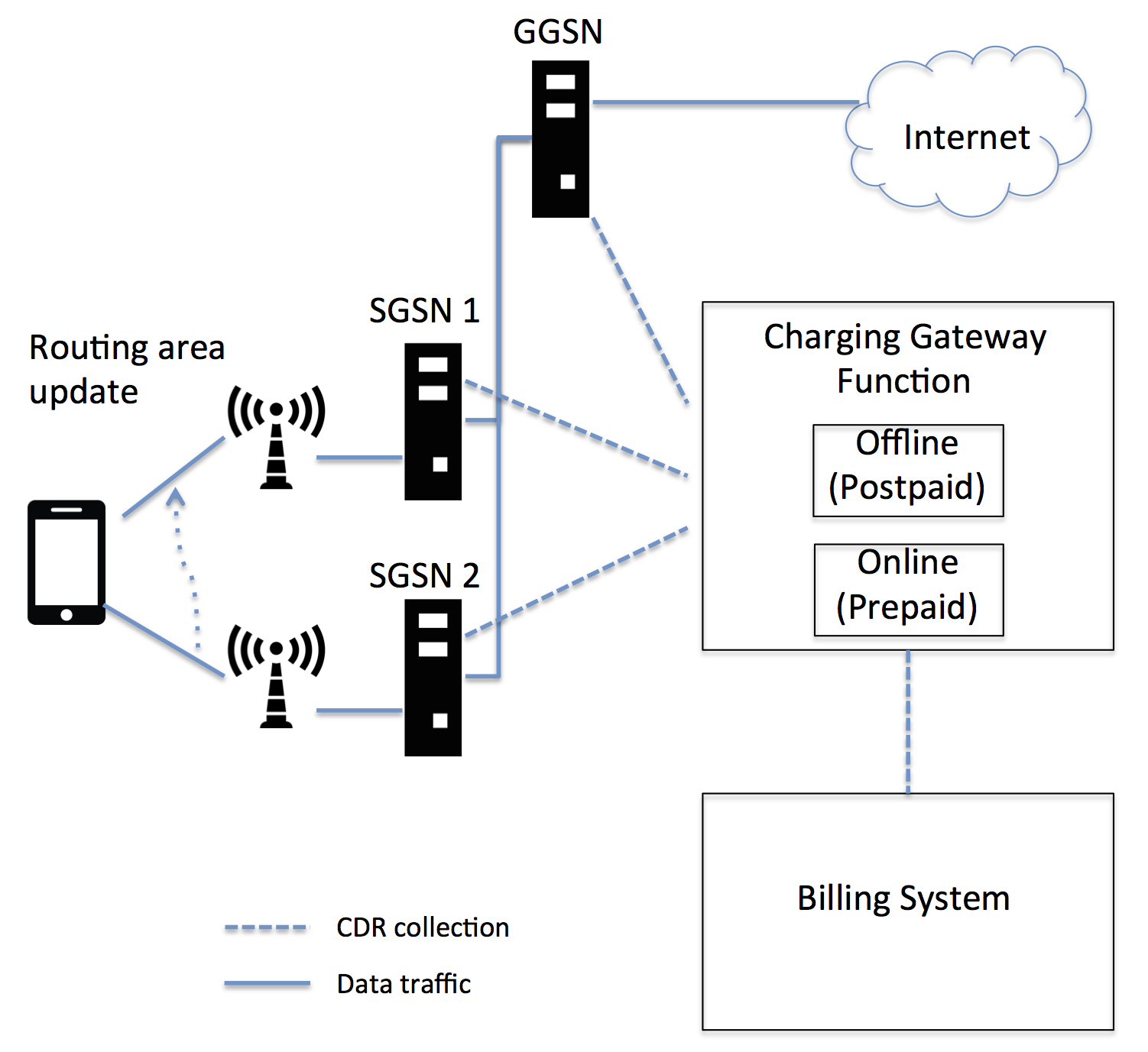}
\caption{Charging-related processes \cite{korhonen2003}. A single subscriber can trigger the generation of CDR in
different SGSN nodes. In order to synchronize the session and register traffic usage, the GGSN node also generates CDR.
The \emph{Billing System} receives all CDR and calculates the costs of the services used by each subscriber.}
\label{fig:charging}
\end{figure}

To avoid part of the previous difficulties, post-mediation internet traffic CDR are a cost-effective alternative to use in urban 
mobility applications (see \cite{Calabrese2014} and \cite{Blondel2015}). These kinds of registers are generated by the \emph{Billing System}, as depicted in Fig.~\ref{fig:charging}. 
An internal core network functionality, called \emph{Charging Gateway function}, transfers charging information from the core network gateway nodes 
to the \emph{Billing System}, which is an external core network component. This system is in charge of calculating the cost of the services used by each subscriber, based on current tariffs.
It is provided by special vendors, often different from well known telecommunication-infrastructure vendors. 
Since this kind of CDR is usually stored in a mid-term persistent storage far from the 
network and operation nodes (typically a data warehouse used by business processes), ETLs can be applied to the dataset without 
further critical monitoring. The remaining problem that must be solved is the time accuracy of the events. 

A methodology to
obtain insights on the distribution of the time accuracy of post-mediation CDR is proposed. We use network traffic events as ground-truth 
in order to obtain timely error measurements for the CDR. Using these error measurements, time-series analysis techniques are applied to
infer similarities in the error behaviour for each \emph{BTS}, \emph{Node} and \emph{eNode} in the service provider's network. 

%
%


\section{Case study}

Our case study is based on anonymized data collected from prepaid and postpaid post-mediation CDR.
The data is provided by Movistar - Telef\'onica Chile, 
with approximately 33\% of the market share \cite{subtel}, with more than 7 million subscribers and 40,000 network cells \cite{antenas}.

\subsection{Error measurement}

As ground-truth we use data provided by Huawei Smartcare SEQ Analyst \cite{smartcare}, 
a network event measurement tool which collects various signaling/protocol data 
from the core network. Particularly, each internet traffic session is measured with its
corresponding initial/final timestamps, byte consumption and cell ID 
of origin, among other information. 
Originally intended to help in quality of service measurement and operations efficiency,
in our case we compare this data against post-mediation CDR. 
We identify the last network event registered in a particular network cell which is closest in time to each
CDR event. The comparison is performed backwards in time, since for each CDR event there is at least one
network event which happened in the near past in the same cell.  
In Table \ref{table_cdr-smart} a sample event trace for an anonymous subscriber is depicted. For each CDR event, a backward search
is performed until a network event is registered in the same cell. 
For instance, 
at time 72198 (20:03:18) the CDR event has an associated network event at time 69406 (19:16:46 at cell A2).
We measure the error for all anonymous subscribers during 5 working days of May 2016, 
aggregating the errors by cell, technology (2G, 3G and 4G) and charging type, using anonymized Cell IDs. 

\begin{table}[!t]
\caption{Error measurement using CDR and network events. The closest past network event in the same cell is associated to each CDR.}
\label{table_cdr-smart}
\centering 
\begin{tabular}{l|l|l|l|l}\hline\hline
Time (s) & CellID/Tech & Type & Charging & Error (s)\\\hline\hline
69405&A1/2G&Network&-&-\\\hline 
69406&A2/4G&Network&-&-\\\hline 
69499&A1/2G&CDR&Prepaid&94\\\hline 
72198&A2/4G&CDR&Postpaid&2792\\\hline\hline 
\end{tabular}
\end{table}

\subsection{Error analysis}

\begin{figure}[t] \centering
  \includegraphics[width=9cm]{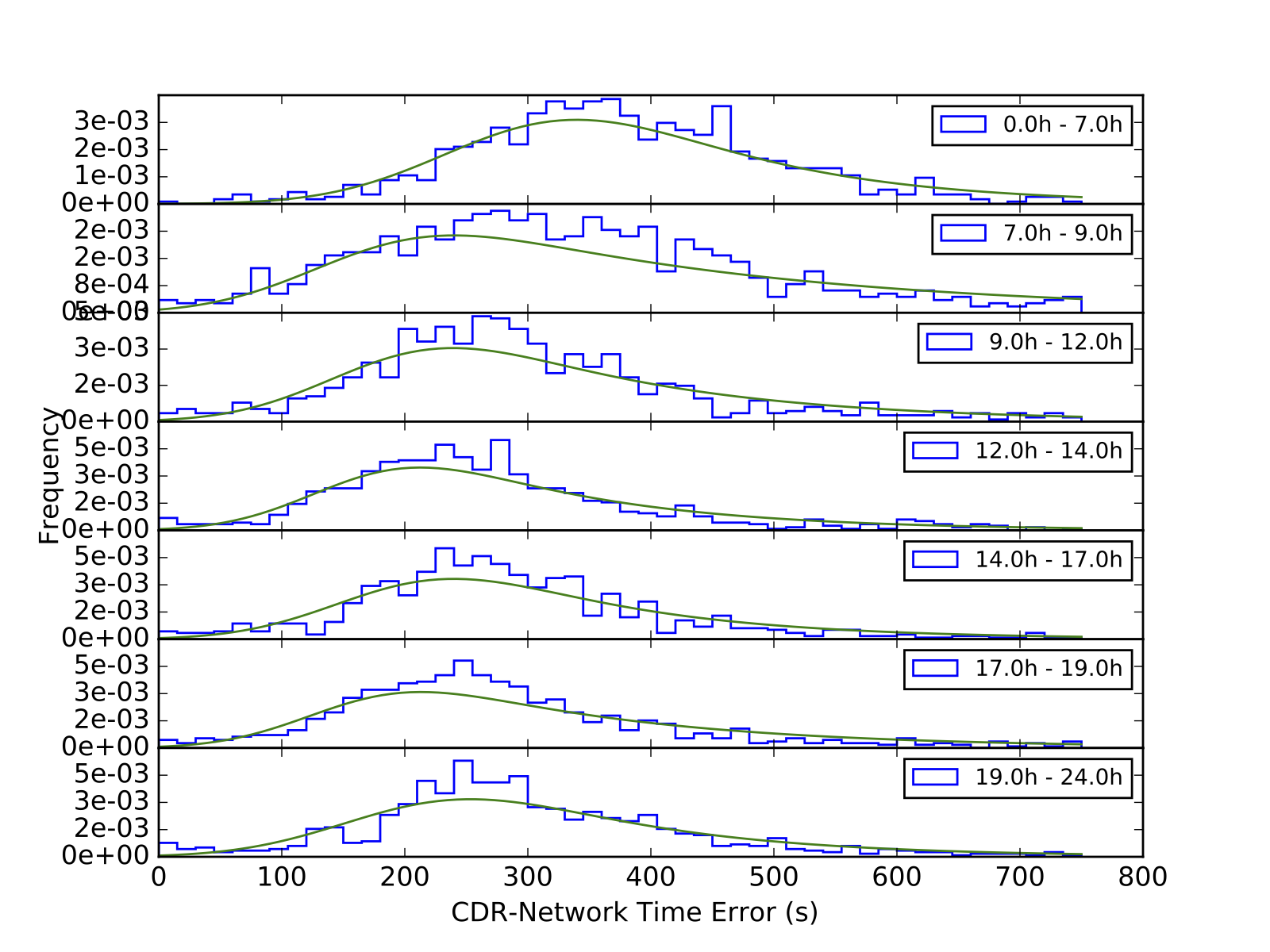}
  \includegraphics[width=9cm]{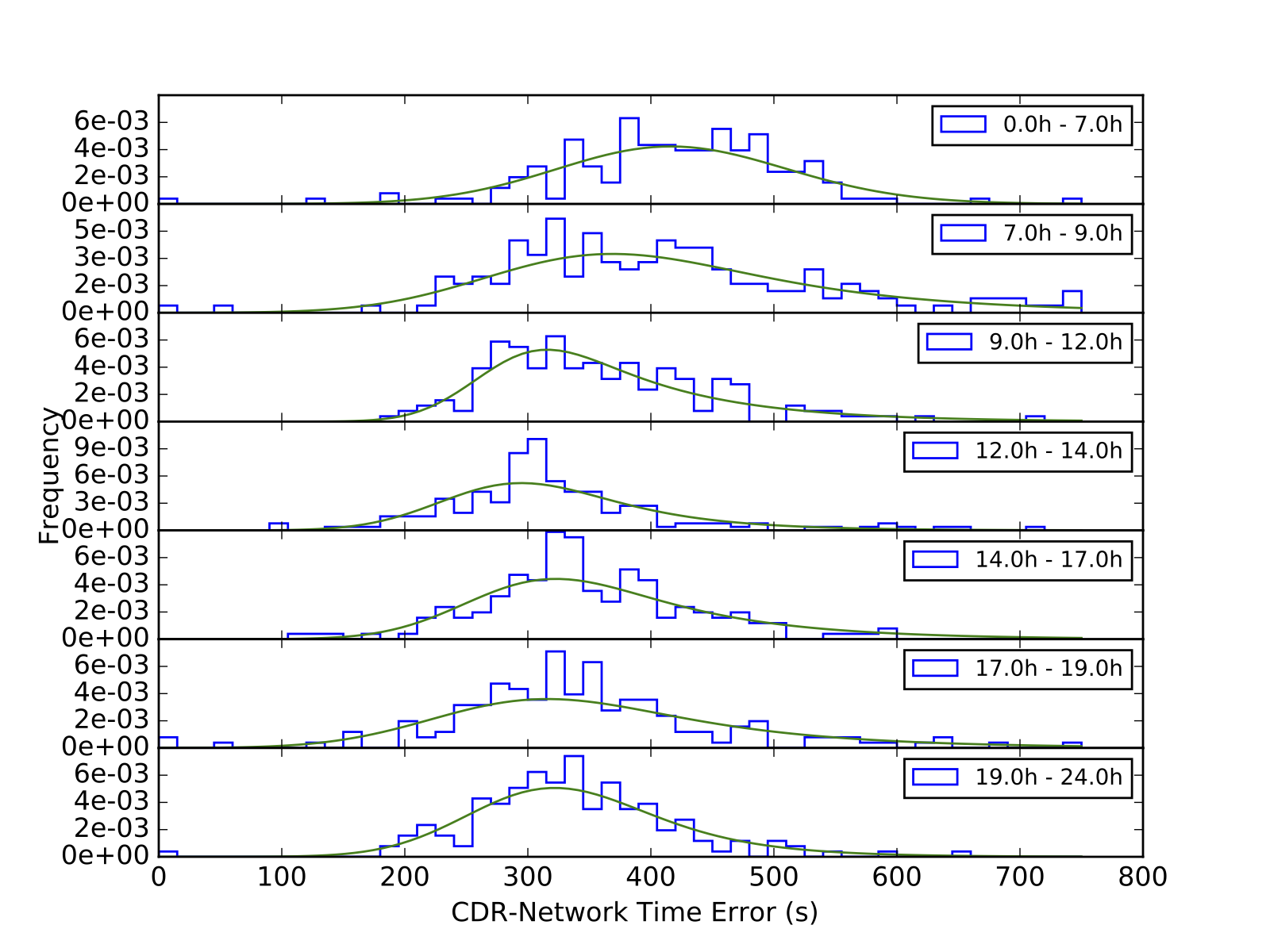}
  \caption{Normed distribution of the CDR-Network time errors for
    different time-bins. Only cells from La Serena-Coquimbo region
    were selected, and the error were separated with respect to
    technologies (2G, 3G - top Panel, 4G - bottom panel) and types of
    contract (prepaid not shown, postpaid). The green line is the best
    fit (exponentially modified normal law).}
  \label{fig:CDR_error_distribution}
\end{figure}

\begin{figure}[t] \centering
  \includegraphics[width=9cm]{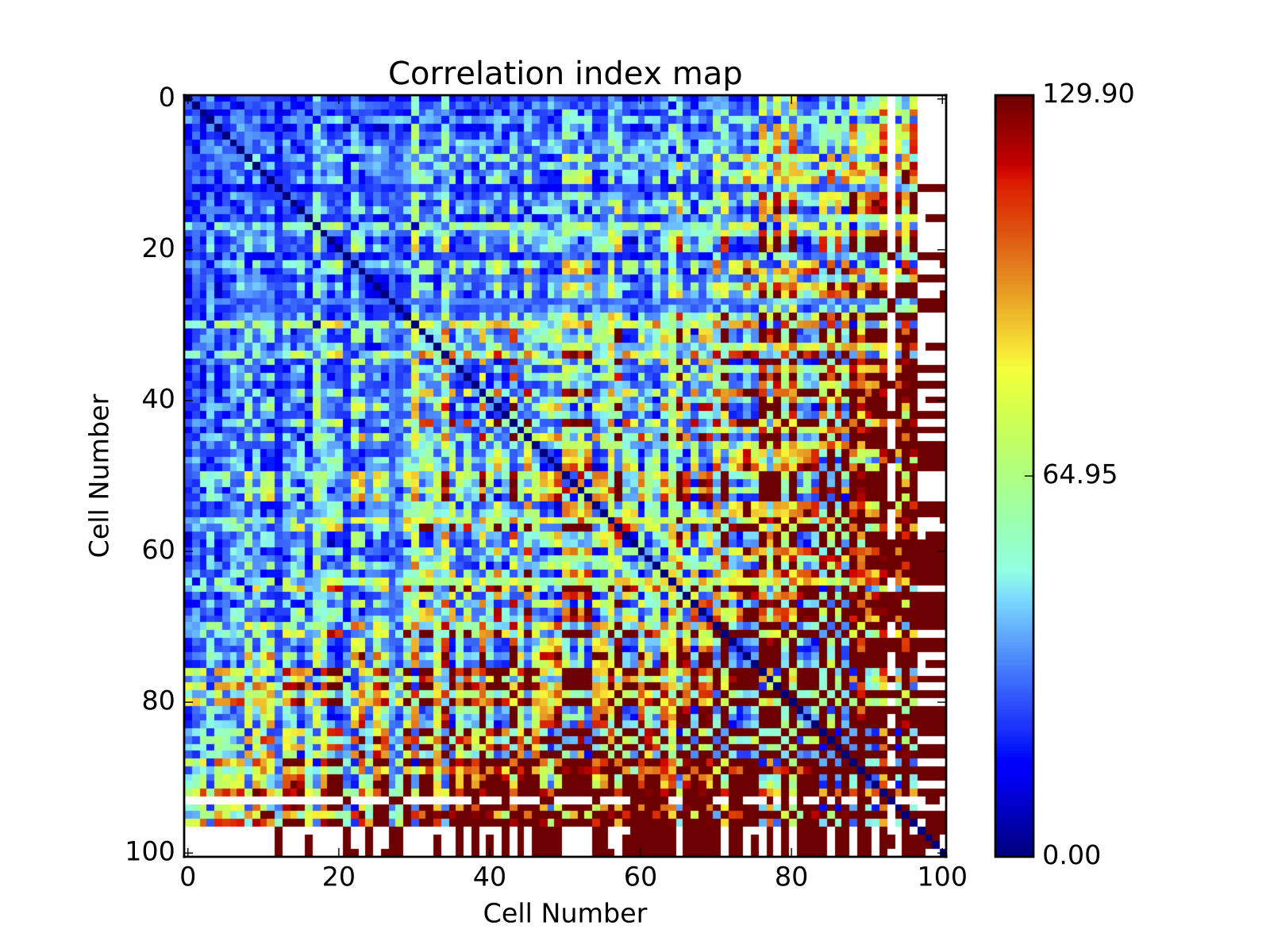}
  \caption{Cell-cell correlation index (as explained in
    Sect.~\ref{subsub:Cell_to_cell_similarities}) matrix of CDR time
    errors for a random sample of 100 antenna cells. The lower the
    index (blue color), the better the correlation.}
  \label{fig:Box_analysis_simi_matrix}
\end{figure}

Various analysis can be performed to quantify and understand the
behavior of the CDR-Network time errors. In this section, we only
focus on the daily variability and the cell to cell
similarities. This allow us to investigate both the temporal and
spatial behaviors.

\subsubsection{Daily variability}
\label{subsub:Daily_variability}

To assess the variability of the CDR-Network time errors along the
day, we split the daily error distributions into time-bins. We then
compute the mean value of time errors for each bin. These two steps
are repeated for the set of antenna cells in the La Serena-Coquimbo
region (500,000 inhabitants - data from May 10$^{\textrm{th}}$; see
Fig.~\ref{fig:CDR_error_distribution}). Both 3G and 4G technologies
distributions vary along the day with error larger at night (bin [0h -
  7h[). The error then progressively diminishes along the day. These
    distributions are best represented by an exponentially modified
    normal law.  We note that the 3G technology induces smaller errors
    that the 4G one (see table~\ref{table:mean_std}). However, the
    scarce 2G data does not allow us to draw conclusions at this
    moment. The same behavior are found for both prepaid (not shown)
    and postpaid.

\begin{table}
\centering
\caption{Mean and Std. Dev. for each bin and each cell technology.}
\label{table:mean_std}
\begin{adjustbox}{width=\columnwidth}
\begin{tabular}{cccccccc}
\cline{3-8}
    \multicolumn{2}{c}{}&\multicolumn{2}{c}{2G}& \multicolumn{2}{c}{3G}&\multicolumn{2}{c}{4G} \\       
\cline{2-8}
\multicolumn{1}{c}{} & time bins &  mean (s)&      std (s)&      mean (s)&     std (s)&      mean (s)&      std (s)\\

\cline{3-8}
\hline
        & [0h - 7h[   &479.17&   273.26&    399.81&  330.84&    419.83&    95.46\\
        & [7h - 9h[   &706.36&  1377.55&    412.66&  560.52&    420.76&   154.17\\
        & [9h - 12h[  &523.53&   791.01&    330.09&  247.35&    364.24&   108.46\\
Prepaid& [12h - 14h[ &501.32&   655.23&    291.51&  215.58&    321.89&    94.69\\
        & [14h - 17h[ &382.50&   307.71&    309.88&  199.30&    366.76&   146.57\\
        & [17h - 19h[ &446.99&   447.46&    320.68&  288.98&    362.92&   154.60\\
        & [19h - 24h[ &551.35&   764.46&    320.39&  196.56&    341.96&    93.65\\
\hline
\hline
        & [0h - 7h[   &477.34&   652.55&    343.86&  866.33&    389.69&   205.59\\
        & [7h - 9h[   &638.48&  2161.95&    474.68&  751.63&    475.75&   480.66\\
        & [9h - 12h[  &477.67&   649.28&    377.38&  345.09&    450.79&   709.37\\
Postpaid& [12h - 14h[ &423.49&   571.16&    364.22&  339.66&    397.20&   380.28\\
        & [14h - 17h[ &691.26&  3523.42&    394.98&  386.74&    382.21&   237.71\\
        & [17h - 19h[ &726.62&  1962.85&    412.13&  540.58&    419.53&   387.26\\
        & [19h - 24h[ &510.77&   615.65&    376.42&  417.74&    543.97&  1229.22\\
\hline
\hline

\end{tabular}
\end{adjustbox}
\end{table}

\subsubsection{Cell to cell similarities}
\label{subsub:Cell_to_cell_similarities}

We used the same time-of-the-day bin splitting to compare the
cells. For each of these bins, we set up the histograms of error. We
then compare the time-bin histograms between cells thanks to a
Kolmogorov-Smirnov test and combine the $p$-values of the tests using
Fisher's method. This results in a correlation index that quantifies
the behavior similarities between two cells. We end up creating the
full matrix of correlation indices as depicted in
Fig.~\ref{fig:Box_analysis_simi_matrix}. For clarity, in this figure,
we only present a random selection of 100 antenna cells. The matrix is
ordered according to the sum of all of the cell-to-cell correlation
indices. Cells having a close behavior are segregated in the top left
of the matrix. This matrix can help to distinguish generic cell
behaviors. For instance, when ordered, we clearly observe a blue
region (same behavior between cells) and a red region (different
behavior between cells). A first generic group combines cell from the
blue region. Moreover, the presence of blue points in the red region
indicates that these exotic cells may form a second group. We finally
note the presence of white points that refer to cells without a large
enough number of events.

\section{Conclusion}
A methodology to analyze time accuracy in post-mediation CDR
is presented. Exploratory analysis shows that different behaviours 
arise when studying disaggregated data. Particularly between postpaid 3G and postpaid 4G,
by analyzing a sample data from a mid-size city.
Further analysis is needed, but the overall objective of the methodology can lead
to accurate error models for this kind of datasets. 


\section*{Acknowledgment}

The authors would like to thank Movistar - Telef\'onica Chile
and Chilean government initiative CORFO 13CEE2-21592 (2013-21592-1-INNOVA\_PRODUCCION2013-21592-1).



\bibliographystyle{IEEEtran}
%
%
%

\end{document}